\documentclass{article} 
\usepackage{iclr2019_conference,times}

\usepackage{amsmath,amsfonts,bm}

\def\eqref#1{equation~\ref{#1}}

\def\1{\bm{1}}

\DeclareMathAlphabet{\mathsfit}{\encodingdefault}{\sfdefault}{m}{sl}
\SetMathAlphabet{\mathsfit}{bold}{\encodingdefault}{\sfdefault}{bx}{n}

\usepackage{hyperref}
\usepackage{url}
\usepackage{graphicx}
\usepackage{caption}
\usepackage{cleveref}
\usepackage{textgreek}
\usepackage{xspace}
\usepackage[binary-units]{siunitx}
\usepackage{amsmath}
\usepackage{booktabs}
\usepackage{listings}
\usepackage{xcolor}
\usepackage{verbatim}

\colorlet{punct}{red!60!black}
\definecolor{background}{HTML}{F3F3F3}
\definecolor{delim}{RGB}{20,105,176}
\colorlet{numb}{magenta!60!black}

\lstdefinelanguage{json}{
    basicstyle=\normalfont\ttfamily,
    numbers=left,
    numberstyle=\scriptsize,
    stepnumber=1,
    numbersep=8pt,
    showstringspaces=false,
    breaklines=true,
    frame=lines,
    backgroundcolor=\color{background},
    literate=
      {:}{{{\color{punct}{:}}}}{1}
      {,}{{{\color{punct}{,}}}}{1}
      {\{}{{{\color{delim}{\{}}}}{1}
      {\}}{{{\color{delim}{\}}}}}{1}
      {[}{{{\color{delim}{[}}}}{1}
      {]}{{{\color{delim}{]}}}}{1},
}

\newcommand{\openarxiv}{arXiv\xspace}
\newcommand{\eat}[1]{}

\title{On the Use of ArXiv as a Dataset}

\iclrfinalcopy

\author{Colin B. Clement \\
Cornell University, Department of Physics\\Ithaca, New York 14853-2501, USA\\
\texttt{cc2285@cornell.edu} \\
\And
Matthew Bierbaum \\ 
Cornell University, Department of Information Science\\Ithaca, New York 14853-2501, USA\\
\texttt{mkb72@cornell.edu} \\
\And 
Kevin O'Keeffe \\
Senseable City Lab, Massachusetts Institute of Technology \\
Cambridge, MA 02139 \\
\texttt{kokeeffe@mit.edu} \\
\And
Alexander A. Alemi \\
Google Research \\
Mountain View, CA \\
\texttt{alemi@google.com}
}

\begin{document}

\maketitle

\begin{abstract}
The arXiv has collected 1.5 million pre-print articles over 28 years, hosting literature from scientific fields including Physics, Mathematics, and Computer Science. Each pre-print features text, figures, authors, citations, categories, and other metadata. These rich, multi-modal features, combined with the natural graph structure---created by citation, affiliation, and co-authorship---makes the arXiv an exciting candidate for benchmarking next-generation models. Here we take the first necessary steps toward this goal, by providing a pipeline which standardizes and simplifies access to the arXiv's publicly available data. We use this pipeline to extract and analyze a 6.7 million edge citation graph, with an 11 billion word corpus of full-text research articles. We present some baseline classification results, and motivate application of more exciting generative graph models.
\end{abstract}

\section{Introduction}
Real world datasets are typically multimodal (comprised of images, text, and time series, etc) and have complex 
relational structures well captured by a graph. Recently, advances have been made on models which act on graphs, allowing the rich features and relational structures of real-word data to be utilized ~\citep{hamilton2017representation,hamilton2017inductive, battaglia2018relational,goyal2018graph,nickel2016review}. Many of these advances have been facilitated by
the availability of large, benchmark datasets: for example, the ImageNet~\citep{ILSVRC15} dataset has been widely used as a community standard for image classification. We believe the arXiv can provide a similarly useful benchmark for large scale, multimodal, relational modelling.

The arXiv\footnote{https://arxiv.org} is the de-facto online manuscript pre-print service for Computer Science, Mathematics, Physics, and many interdisciplinary communities.
Since 1991 the arXiv has offered a place for researchers to reliably share their work as it undergoes the process of peer-review, and for many researchers it is their primary source of literature.
With over 1.5 million articles, a large multigraph dataset can be built, including full-text articles, article metadata, and internal co-citations.

The arXiv has been used many times as a dataset. \citet{liben2007link} used the topology of the arXiv co-authorship graph to study link prediction. \citet{dempsey2019hierarchical} used the authorship graph to test a hierarchically structured network model.
 \citet{DBLP:journals/corr/LopuszynskiB13} used the category labels of arXiv documents to train and assess an automatic text labelling system. \citet{DBLP:journals/corr/DaiOL15} used a subset of the full text available on the arXiv to study the utility of ``paragraph vectors'' for capturing document similarity. \citet{DBLP:journals/corr/AlemiG15} used the fulltext to evaluate a method for unsupervised text segmentation. \citet{eger2019predicting} and
\citet{liu2018using} built models to predict future research topic trends in machine learning and physics respectively. The arXiv also formed the basis of the popular 2003 KDD Cup~\citep{gehrke2003overview}, in which researchers competed for the prize of best algorithm for citation prediction, download estimation, and data cleaning\footnote{The data for those challenges are available at \url{http://www.cs.cornell.edu/projects/kddcup/datasets.html}}. 

All these works used different subsets of arXiv's data, limiting their potential impact, as future researchers will be unable to directly compare their work to these existing results. The goal of this paper is to improve this situation by providing an open-source pipeline to standardize, simplify, and normalize access to the arXiv's public data, providing a benchmark to facilitate the development of models for multi-modal, relational data.




\section{Dataset}

We built a freely available, open-source pipeline\footnote{\url{https://github.com/mattbierbaum/arxiv-public-datasets/releases/tag/v0.2.0}}
for collecting arXiv metadata from the Open Archive Initiative~\citep{lagoze2001open}, and bulk \textsc{pdf} downloading from the arXiv\footnote{\url{https://arxiv.org/help/bulk_data}}. Further, this pipeline converts the raw \textsc{pdf}s to plaintext, builds the intra-arXiv co-citation network by searching the full-text for arXiv \texttt{id}s, and cleans and normalizes \texttt{author} strings.

\subsection{Metadata}

Through its participation in the Open Archives Initiative,\footnote{\url{http://www.openarchives.org/}} the arXiv makes all article metadata\footnote{\url{https://arxiv.org/help/prep}} available, with updates made shortly after new articles are published\footnote{Further details available at \url{https://arxiv.org/help/oa}}. We provide code for utilizing these public
APIs to download a full set of current arXiv metadata. As of 2019-03-01, metadata for 1,506,500 articles was available.
For verification and ease of use purposes, we provide 
a copy of the metadata (less abstracts) on the date
we accessed it.
An example listing is shown in~\Cref{fig:metadata}. Each article includes an arXiv id (e.g. \texttt{0704.0001})\footnote{There are two forms of valid arXiv IDs, delineated by the year 2007, described in \url{https://arxiv.org/help/arxiv_identifier}.} 
used to identify the article, the publicly visible name of the submitter, a list of authors, title, abstract, versions
and category listings, as well as optional \texttt{doi}, \texttt{journal-ref} and \texttt{report-no} fields.
Of particular note is the first category listed, the \emph{primary} category, of which there are 171 at this time. Notice that the list of authors is just a single string of author names, potentially joined with commas or `and's.  We've
 provided a suggested normalization and splitting script for splitting these
 \texttt{authors} strings into a list of author names. Additional fields may be present to denote \texttt{doi}, \texttt{journal-ref} and \texttt{report-no}, although these are not validated they can potentially be used to find intersections between the arXiv dataset and other scientific literature datasets. Population counts for the optional fields are shown 
 in~\Cref{tab:counts}.

\begin{figure}[htbp]
\centering
\tiny
\begin{lstlisting}[language=json,firstnumber=1]
{'id': '1904.99999',
 'submitter': 'Colin B. Clement',
 'authors': 'Colin B. Clement, Matthew Bierbaum, Kevin P. O\'Keeffe, and Alexander A. Alemi',
 'title': 'On the Use of ArXiv as a Dataset',
 'comments': '7 pages, 3 figures, 2 tables',
 'journal-ref': '',
 'doi': '',
 'abstract': 'The arXiv has collected 1.5 million pre-prints over 28 years, hosting literature from physics, mathematics, computer science, biology, finance, statistics, electrical engineering, and economics. Each pre-print features text, figures, author lists, citation lists, categories, and other metadata. These rich, multi-modal features, combined with the natural relational graph structure created by citation, affiliation, and co-authorship makes the arXiv an exciting candidate for benchmarking next-generation models. Here we take the first necessary steps toward this goal, by providing a pipeline which standardizes and simplifies access to the arXiv's publicly available data. We use this pipeline to extract and analyze a 6.7 million edge citation graph, with an 11 billion word corpus of full-text research articles. We present some baseline classification results, and motivate application of more exciting relational neural network models.'
 'categories': ['cs.IR'],
 'versions': ['v1']}
\end{lstlisting}
\caption{\label{fig:metadata}An example of what the metadata for this very article may look like if it were submitted to the arXiv.}
\end{figure}

\begin{table}[htbp]
    \centering
    \begin{tabular}{rcccccc}
         Count & 1,506,500 & 1,491,303 & 1,229,138 & 810,209 & 608,286 & 154,922 \\ \midrule
         Field & \texttt{id} & \texttt{submitter} & \texttt{comments} &  
         \texttt{doi} & \texttt{journal-ref} & \texttt{report-no} 
    \end{tabular}
    \caption{Number of articles with the corresponding field populated.  Note that the fields \texttt{id}, \texttt{abstract}, \texttt{authors}, \texttt{versions}, and \texttt{categories} are always populated.}
    \label{tab:counts}
\end{table}

\subsection{Full Text}
One advantage the arXiv has over other graph datasets is that it provides a very rich attribute at each \texttt{id} node: the full raw text and figures of a research article. To extract the raw text from \textsc{PDF}s, we provide a pipeline with two parts. A helper script downloads the full set of \textsc{pdf}s available through the arXiv's bulk download service\footnote{\url{https://arxiv.org/help/bulk_data}}. Since arXiv hosts their data in a requester-pay \textsc{AWS} \textsc{S3} buckets, this constitutes $\sim \SI{1.1}{\tera\byte}$ and $\sim\$100$ to fully download.  
For posterity, we have provided \textsc{MD5} hashes of the \textsc{pdf}s at the state of the frozen metagraph extraction.
Raw \TeX\ source is also available for the subset of articles that provide it. 
Second, we provide a standard \textsc{pdf}-to-text converter -- powered by \texttt{pdftotext}\footnote{Version 0.61.1, available on most Debian systems from the \texttt{apt} package \texttt{poppler-utils}} --  to convert the \textsc{pdf}s to plaintext. 

Using this pipeline, it is currently possible to extract a corpus of 1.37 million raw text documents. \Cref{fig:example_text} shows an example 
of the text extracted from a \textsc{pdf}.
Though the extracted text isn't perfectly clean, we believe it will still be useful for many tasks, and hope future contributions to our repository will provide better data cleaning procedures. 


The extracted raw-text dataset is $\sim \SI{64}{\giga\byte}$ in size, totaling $\sim 11$ billion words. An order of magnitude larger than the common billion word corpus~\citep{chelba2013one}, this large size makes the arXiv raw-text a competitive alternative to other full text datasets. Moreover, the technical nature of the arXiv distinguishes it from other full text datasets. For example, the \TeX\ data contained in the arXiv presents an opportunity to study mathematical formulae in bulk, as is done in the NTCIR-11 Task: Math-2~\citep{aizawa2014ntcir}.

\begin{figure}[htbp]
    \centering
    \tiny
\begin{lstlisting}[language=json,firstnumber=1]
Published as a conference paper at ICLR 2019

O N THE U SE OF A R X IV AS A DATASET
Colin B. Clement
Cornell University, Department of Physics
Ithaca, New York 14853-2501, USA
cc2285@cornell.edu

Matthew Bierbaum
Cornell University, Department of Information Science
Ithaca, New York 14853-2501, USA
mkb72@cornell.edu

Kevin O’Keeffe
Senseable City Lab, Massachusetts Institute of Technology
Cambridge, MA 02139
kokeeffe@mit.edu

Alexander A. Alemi
Google Research
Mountain View, CA
alemi@google.com

A BSTRACT
The arXiv has collected 1.5 million pre-print articles over 28 years, hosting literature from scientific fields including Physics, Mathematics, and Computer Science. Each pre-print features text, figures, authors, citations, categories, and other
metadata. These rich, multi-modal features, combined with the natural graph
structure—created by citation, affiliation, and co-authorship—makes the arXiv
an exciting candidate for benchmarking next-generation models. Here we take the
first necessary steps toward this goal, by providing a pipeline which standardizes
and simplifies access to the arXiv’s publicly available data. We use this pipeline to
extract and analyze a 6.7 million edge citation graph, with an 11 billion word corpus of full-text research articles. We present some baseline classification results,
and motivate application of more exciting generative graph models.
\end{lstlisting}
    \caption{\label{fig:example_text}
    Example text extracted from this \textsc{pdf}. 
    }
\end{figure}

\subsection{Co-Citations}
While the arXiv does not currently publicly provide an \textsc{api} to 
access co-citations, our pipeline allows a simple but large co-citation network to be extracted. We extracted this network by searching the text of each article for valid arXiv \text{id}s, thereby finding which nodes should be linked to a given node in the co-citation network. We provide a compressed binary of the resulting network at the repository\footnote{As part of one of the tagged releases: \url{https://github.com/mattbierbaum/arxiv-public-datasets/releases}}, so that researchers can study it directly, and avoid the difficulty of constructing it themselves. \Cref{tab:citation_stats} summarizes the size and statistical structure of our co-citation network, compared with other popular citation networks. \citet{vsubelj2014network} also studied data from the arXiv, but as indicated in the bottom row of~\Cref{tab:citation_stats}, it used only the 34,546 articles from the 2003 KDD Cup challenge.

\Cref{tab:citation_stats} reports standard statistics for the co-citation network. Our \openarxiv co-citation network contains $O(10^6)$ nodes, an order of magnitude larger than the $O(10^5)$ nodes in the other citation networks. The exponents of best fit for the degree distributions $\alpha_\mathrm{in}$ and $\alpha_\mathrm{out}$ are consistent with the existing citation networks~\citet{vsubelj2014network}, as it the the degree $\langle k \rangle$. $62 \%$ of the nodes are contained in the largest weakly connected component, while $31 \%$ of the nodes are fully isolated -- meaning their in-degree $k_\mathrm{in}$ and out-degree $k_\mathrm{out}$ are zero. Recall that our \openarxiv co-citation network only contains publications which have been posted on the arXiv; a given paper which cites papers published elsewhere -- and not on the arXiv -- will have $k_{out} = 0$ in this set, which is an explanation the large number of isolated nodes.

\begin{table}[t]
\begin{center}
    \caption{\label{tab:citation_stats}\textbf{Graph statistics for popular citation networks}. All but the data for this work (first row) were taken from Table 1 and 2 in \cite{vsubelj2014network}. $ \langle k \rangle$ is the average degree, and $\alpha_\mathrm{in}$ and $\alpha_\mathrm{out}$ are power law exponents of best fit for the degree distribution. WCC refers to the largest weakly connected components, computed using the python package `networkx'.
    The power law exponents $\alpha_\mathrm{in}, \alpha_\mathrm{out}$ were found using the python module \texttt{powerlaw}. When fitting data to a powerlaw, the package discards all data below an automatically computed threshold $x_\mathrm{min}$. These thresholds for $k_\mathrm{in}$ and $k_\mathrm{out}$ were $x_\mathrm{min} = 73$ and $x_\mathrm{min} = 59$ respectively.}
\begin{tabular}{ lSSSSSS }
\toprule
Dataset & {$N_\mathrm{nodes}$} & {$N_\mathrm{edges}$} & {$\langle k \rangle$} & {$\alpha_\mathrm{in}$} & {$\alpha_\mathrm{out}$} & {\% WCC}\\
\midrule
{\bf \openarxiv} & \num{1.35e6} & \num{6.72e6} & 9.933 & 2.93 & 3.93 & 62 \\
WoS & \num{1.40e5} & \num{6.4e5} & 9.11 & 2.39 & 3.88 & 97 \\
CiteSeer & \num{3.84e5} & \num{1.74e6} & 9.08 & 2.28 & 3.82 & 95 \\
KDD2003 & \num{3.34e4} & \num{4.21e5} & 24.50 & 2.54 & 3.45 & 99.6 \\
\bottomrule
\end{tabular}
\end{center}
\end{table}


%

Beyond constructing and analyzing a co-citation network, the arXiv dataset can be used for many tasks, such as relationally powered classification, author attribution, segmentation, clustering, structured prediction, language modeling, link prediction and automatic summary generation. As a basic demonstration, in \Cref{tab:classification} we show some baseline category classification results. These were obtained by training logistic regression on 1.2 million arXiv articles to predict in which category (e.g. \texttt{cs.Lg}, \texttt{stat.ML}) a given article resides. See Appendix~\ref{appendix:lr} for a detailed explanation of the experimental setup. Titles and abstracts were represented by vectors from a pre-trained instance\footnote{From \url{https://tfhub.dev/google/universal-sentence-encoder/2}} of the Universal Sentence Encoder of \citet{universalsentence}. We see that including more aspects of each document (titles, abstracts, fulltext) and exposing their relations via co-citation leads to better predictive power. This is only scratching the surface of possible tasks and models applied to this rich dataset.

\begin{table}[htbp]
    \centering
        \caption{Baseline classification performance on a holdout set of 390k articles. Titles and abstracts
    were embedded in a 512 dimensional subspace using the Universal Sentence Encoder, and trained on 1.2 million articles with
    logistic regression. `All' refers to the concatenation of titles, abstract, fulltext, and co-citation features. `All - X' refers to the ablation of feature X from `All.' Top $n$ is the classification accuracy testing when the correct class is in the top $n$ most confident predictions. Detailed explanation of the features and methods can be found in Appendix~\ref{appendix:lr}.}
    \begin{tabular}{lrrrr}
    \toprule
         Features & Top 1 & Top 3 & Top 5 & Perplexity \\
         \midrule
         Titles (T) & 36.6\% & 59.3\% & 68.8\% & 12.7 \\
         Abstracts (A) & 46.0\% & 70.7\% & 79.5\% & 7.5 \\
         Fulltext (F) & 64.2\% & 79.4\% & 85.9\% & 4.6 \\
         Co-citation (C) & 37.8\% & 49.4\% & 53.8\% & 18.5\\
         {\bf All = T + A + F + C} & {\bf 78.4\%} & {\bf 91.4\%} & {\bf 94.5\%} & {\bf 2.3}\\
         All - T & 77.0\% & 90.7\% & 94.0\% & 2.5\\
         All - A & 74.7\% & 88.3\% & 91.9\% & 2.8\\
         All - F & 59.0\% & 79.8\% & 86.2\% & 4.6\\
         All - C & 75.5\% & 89.9\% & 93.6\% & 2.6\\
         \bottomrule
    \end{tabular}
    \label{tab:classification}
\end{table}

\section{Conclusion}

As research moves increasingly towards structured relational modelling~\citep{hamilton2017representation,hamilton2017inductive,battaglia2018relational}, there is a growing need for large-scale, relational datasets with rich annotations. With its authorship, categories, abstracts, co-citations, and full text, the arXiv presents an exciting opportunity to promote progress in relational modelling. We have provided an open-source repository of tools that make it easy to download and standardize the data available from the arXiv. Our preliminary classification baselines support the claim that each mode of the arXiv's feature set allows for greatly improved category inference. More sophisticated models that include relational inductive biases---encoding the graph structures of the arXiv---will improve these results. Further, this new benchmark dataset will allow more rapid progress in tasks such as link prediction, automatic summary generation, text segmentation, and time-varying topic modeling of scientific disciplines.

\section*{Acknowledgements}
The authors thank the anonymous reviewers for their helpful comments. CBC was funded by NSF grant DMR-1719490. MB thanks the Allen Institute for Artificial Intelligence for funding. KPO thanks the members of the MIT Senseable City Lab consortium for their support.

\bibliography{arxivnet}
\bibliographystyle{iclr2019_conference}







\appendix

\section{Logistic Regression Article Classification Baseline}
\label{appendix:lr}

ArXiv articles are assigned primary categories (e.g. \texttt{cs.AI} is Artifical Intelligene and \texttt{cs.CC} is computational complexity) by the article submitter, which is then confirmed by the ArXiv moderation system. This label can be obtained for each article from the OAI metadata described in the main article, and is the first element of a space-delimited string in the \texttt{categories} attribute. There are, at the time of writing, $L=175$ possible categories. Since more categories can be added in the future and the metadata can be modified, please consult the frozen metadata file in the github repository release\footnote{\url{https://github.com/mattbierbaum/arxiv-public-datasets/releases}} for these 175 categories. This appendix explains how we developed the article classification baselines using features from the titles, abstracts, full-text, and co-citation network. The code for performing this task can be found in the \texttt{git} repository\footnote{\url{https://github.com/mattbierbaum/arxiv-public-datasets/blob/v0.2.0/analysis/classification.py}}.

\subsection{Building Features}
The title, abstract, and full-text of each article is a variable-length string, and each article has both a title and abstract from the OAI metadata, but not all articles have a full-text \textsc{pdf}. In our frozen dataset there are $N=1,506,500$ articles with metadata, but only 1,357,536 have full-text in the ArXiv. We vectorized each string into 512 dimensions using the pretrained Universal Sentence Encoder,\footnote{\url{https://tfhub.dev/google/universal-sentence-encoder/2}} substituting zeros for missing full-text.

The intra-ArXiv citation graph can be used via the $N\times N$ co-citation matrix, which is defined as
\begin{equation}
    M_{ij} = 
    \begin{cases}
        1 \text{~if article~} i \text{~cites article~} j
        \text{~or vice-versa}\\
        0 \text{~else}.
    \end{cases}
\end{equation}
In order to prevent a leaking of the test set into the training set, using the train/test partition defined below,
we omitted citations in $M$ from articles in the training set which connect to the test set, but retained citations in the test set which connect to the training set.

We can also define the $N\times L$ category matrix in the standard one-hot fashion
\begin{equation}
    C_{jl} = 
    \begin{cases}
        1 \text{~if article j is in category~} l\\
        0 \text{~else}.
    \end{cases}
\end{equation}
Then the co-citation feature matrix is the $N\times L$ matrix product $MC$. Note that this feature uses only nearest-neighbor citation graph relationships. We could include next-nearest neighbor relationships and so on by calculating $MC + a M^2C + b M^3 C + \ldots$ for some constants $a$ and $b$. In this paper we only used first order connections via $MC$ as the co-citation feature vectors.

\subsection{Training}

Using vector embeddings from titles, abstracts, and full-text, and co-citation features as described above, we fed several combinations of these vectors concatenated in the obvious way into the \texttt{scikit-learn} SGD classifier
\path{sklearn.linear_model.SGDClassifier}. We used
the keyword arguments \verb|loss='log'|, \verb|tol=1e-6|, \verb|max_iter=50|, and \verb|alpha=1e-7| to define the model, which uses 50 epochs, and very small quadratic regularization \texttt{alpha} on the weights and biases. 

With the features and model defined, we performed a train/test split by shuffling the data in place randomly, and selecting the first $N_\mathrm{train}=1,200,000$ for training. The remaining $N_\mathrm{test}=306,500$ articles were used to evaluate the accuracy of the trained classification, and the model perplexity as reported in table in the main text.


\end{document}